\documentclass[preprint,3p,number,review,sort&compress]{elsarticle}

\usepackage{graphicx}
\usepackage{epstopdf}
\usepackage{psfrag}
\usepackage{amsmath,amssymb}
\usepackage{color}
\usepackage{nicefrac}
\newcommand{\w}{\omega}

\begin{document}


\title{Phase demodulation 
with iterative Hilbert transform embeddings}

\author[up]{Erik Gengel}
 \ead{egengel@uni-potsdam.de}
\author[up,unn]{Arkady Pikovsky}%
\ead{pikovsky@uni-potsdam.de}
\address[up]{%
 Institute for Physics and Astronomy,  University of Potsdam, 
 Karl-Liebknecht Str. 24/25,
 14476 Potsdam, Germany 
}%
\address[unn]{Department of Control Theory,  Institute of Information Technologies,
Mathematics and Mechanics, Lobachevsky University Nizhny Novgorod, Russia}



\date{\today}

\begin{abstract}
We propose an efficient method for demodulation of phase modulated signals via 
iterated Hilbert transform embeddings. We show that while  a usual approach 
based on one application of the Hilbert transform provides only an approximation 
to a proper phase, with iterations the accuracy is essentially improved,
up to precision limited mainly by discretization effects.
We demonstrate that the method is applicable to arbitrarily complex waveforms, 
and to modulations fast compared to the basic frequency.  Furthermore, we 
develop a perturbative theory applicable to a simple cosine waveform, showing 
convergence of the technique.

\end{abstract}


\maketitle

\section{Introduction}

Demodulation of nearly oscillatory signals is a standard problem in signal processing.
In classical communication schemes, one uses amplitude, phase, and frequency modulations
to store information in an analog signal. At the receiving side, demodulation is needed to
extract this information. In a more general context, when one tries to characterize 
the dynamics of an oscillatory
system based on some observable, modulation carries important information
about the system operation. For example, the electrocardiogram is an important signal
characterizing the human cardio-vascular system. One of the widely used tools is demodulation
of this signal via extracting information about the instantaneous periods (RR intervals), with
further analysis of demodulated 
heart rate variability~\cite{Billman-11}. In many situations one considers
a demodulation task for harmonic carrying waveforms only~(see, e.g., \cite{Fitz-07}), but in fact
one quite often needs to demodulate signals with unknown, potentially rather complex
waveform (like the electrocardiogram signal).

In this paper we focus on the problem of demodulation of phase modulated signals
with arbitrary waveforms and arbitrary phase modulations. This task is essential 
for characterization of oscillatory processes resulting from forced and/or coupled 
self-sustained oscillators, e.g. for  synchronization 
analysis~\cite{ rosenblum2001phase,le2001comparison,Quiroga-Kraskov-Kreuz-Grassberger-02}
and for reconstruction of the phase 
dynamics~\cite{kralemann2008phase,English_etal-15,Stankovski_etal-17}. In particular,
in Refs.~\cite{Kralemann_etal-13,topccu2018disentangling} a continuous phase was extracted from the electrocardiogram
signal, and further used to build a phase model for
the cardio-respiratory interaction.

The Hilbert transform (HT) is a widely used tool in oscillatory data analysis and signal processing. 
It lies at the heart of  the analytic signal 
approach \cite{gabor1946theory,bedrosian1962fm,Vakman-96,feldman2011hilbert,Iatsenko_etal-15},
which provides an effective processing of nearly harmonic signals. 
The Hilbert transform is also extensively used for the phase demodulation
in synchronization studies~\cite{pikovsky2001synchronization,rosenblum2001phase}, see also
recent review~\cite{Sun-Li-Tong-12}.
One uses it to perform a two-dimensional embedding of the observed signal; this embedding then
is usually treated as an analytic signal and the phase is extracted as the argument of the complex state
(or, more generally, as an angle of rotation with respect to some point on the embedding plane).
This approach gives a good approximation, but not the exact phase demodulation. 
The quality of the analytic signal based phase demodulation is better for slow modulation, although also in this situation, as will be
illustrated below, it is not exact.

One can attribute non-exactness of the HT embedding to the fact, that
the HT generally mixes phase and amplitude modulations: if one starts with
a purely phase modulated signal, then the HT embedded signal will have amplitude
modulation as well. In signal processing, workarounds exist 
\citep{kuklik2015reconstruction, feldman2011hilbert} that introduce filtering before  application 
of the HT and/or decomposition of the signal into several functions to avoid the mixing. 

In this paper we propose a phase demodulation method based on iterative HT embeddings.
The first step is the same as in the traditional approach, it yields an approximation, but the signal is not
fully demodulated. We use the phase obtained in the first step as a new time variable to perform
the next embedding, with better demodulation. The repetition of this procedure provides better and better
approximations.
We will demonstrate that the iterations result in a  very
accurate demodulation with errors related to numerical errors in calculating the Hilbert  transform integral.
Noteworthy, our method is not restricted to simple waveforms like cosine function; we will
present its implementation for arbitrary (but smooth) waveforms. Thus, no filtering is
involved in the approach we describe.

For completeness, we mention another class of approaches to phase demodulation,
where not the phase itself is determnined, but the instanteneous frequency is evaluated via
a time-frequency decomposition (see, e.g., \cite{Aviyente-Mutlu-11}).
Recently, a variant of this approach -- wavelet-based synchrosqueezing -- have
been suggested~\cite{Daubechies_etal-11,Iatsenko_etal-15,Iatsenko_etal-16}. Comparison with these
methods is a subject of future research.
 
 The paper is organized as follows. First,  in section \ref{sec: pms}, we elaborate on 
our definition of the waveform and the phase for a signal, and 
discuss connections to dynamical systems analysis. 
Then we present our method in sections \ref{sec:ppe},\ref{sec:ihe}. 
Analytic results on the convergence of the method,  based on a
perturbation analysis,  are described in section \ref{sec:theory}. Numerical results,
confirming accuracy of the technique for a range of strongly modulated signals and
 for complex waveforms, 
are presented in section \ref{sec:numtests}. In section~\ref{sec:prot},
a transformation from a phase variable,
obtained in the demodulation procedure, to another phase is discussed.
We end with conclusions in section~\ref{sec:concl}.

\section{Phase modulated signals} \label{sec: pms}
\subsection{Definition of a phase modulated signal} \label{sec: pms def pm signals}
The subject of our study are phase-modulated signals.  
First we define a  \textit{waveform} $S(\varphi)=S(\varphi+2\pi)$ as a $2\pi$-periodic function
of its argument. The argument, interpreted as a \textit{phase}, is a monotonous
function of time $\dot \varphi >0$.  If $\dot\varphi\neq \text{const}$, one speaks of a \textit{phase modulated signal}
\begin{equation}
x(t)=S(\varphi(t))\;. 
\label{eq:pmsig}
\end{equation}
Our task is, given the signal $x(t)$, 
to reconstruct the phase $\varphi(t)$ and the waveform $S(\varphi)$. 

Our method below
is not restricted to simple waveforms: $S(\varphi)$ can be a rather complex function with several maxima 
and minima. Below in the numerical examples and in theoretical considerations we will use
\begin{equation}
S_1(\varphi)=\cos(\varphi)
\label{eq:swf}
\end{equation}
as a simple waveform, and 
\begin{equation}
S_2(\varphi)=\cos(\varphi)-0.7\sin (2\varphi)+\cos (3\varphi)
\label{eq:cwf}
\end{equation}
as a complex waveform.

\subsection{Phase modulation in driven oscillators}
\label{sec:pmdo}
The motivation for this setup comes from the inverse problems for 
nonlinear dynamical systems. Phase modulated 
signals naturally appear if a dynamical system with stable self-sustained oscillations (i.e. with a limit cycle)
is driven by a weak external force 
(or interacts with another dynamical 
system)~\cite{Winfree-80,Kuramoto-84,pikovsky2001synchronization,Ermentrout-Terman-10},
as we briefly outline below.

Suppose we have an autonomous dynamical system
\[
\dot{\vec{y}}=\vec{F}(\vec{y})
\]
with a stable limit cycle $\vec{y}_0(t)$. This means that the system generates periodic oscillations
and a generic scalar observable $x(\vec{y})$ is a periodic function of time $x(\vec{y}_0(t))$. 
On the limit cycle and in its
basin of attraction one can introduce the phase variable $\varphi=\Phi(\vec{y})$, such that 
the limit cycle is parametrized by this phase
$\vec{y}_0(\varphi+2\pi)=\vec{y}_0(\varphi)$, and the phase grows linearly in
time with the natural frequency
of the limit cycle $\dot{\varphi}=\omega$.

If there is a weak external force acting on the system
\[
\dot{\vec{y}}=\vec{F}(\vec{y})+\varepsilon \vec{P}(\vec{y},t),
\]
then, in the first approximation in $\varepsilon$,
a phase-modulated dynamics on the limit cycle represents the full 
oscillation of $\vec{y}(t)$ close to $\vec{y}_0(\varphi(t))$, where the phase obeys the equation
\begin{equation}
\dot\varphi=\omega+\varepsilon Q(\varphi,t)
\label{eq:dynph}
\end{equation}
and $Q=\text{grad}\Phi|_{\vec{y}_0(\varphi)}\cdot  \vec{P}(\vec{y}_0(\varphi),t)$ is determined
by the phase response curve of the limit cycle and the force $\vec{P}$ \citep{brown2004phase}. Accordingly,
the scalar observable $x$ is the phase-modulated signal as described above:
\[
x(t)=x(\vec{y}_0(\varphi(t)))\;.
\]
Here the waveform is determined by the form of the limit cycle and of the scalar observable
$S(\varphi)=x(\vec{y}_0(\varphi))$. The inverse problem in the context of the
theory above is formulated as follows: From scalar observations $x(t)$ 
of a driven dynamical system, one seeks to reconstruct the dynamics of the phase $\varphi(t)$. The
first step here is the reconstruction of the phase $\varphi(t)$, what is the topic of the present study. 

\subsection{Phase vs protophase}
\label{sec:ppvp}

In this section we discuss whether the decomposition of a signal \eqref{eq:pmsig}
into the phase and the waveform is unique. The answer is obvious: No. Indeed, one can perform
a transformation 
\begin{equation}
\psi=\Psi(\varphi),\qquad \Psi(s+2\pi)=\Psi(s)+2\pi,\qquad \Psi'>0,
\label{eq:trp}
\end{equation}
to a new phase variable $\psi$, in terms of which the signal is represented as
\[
x(t)=\tilde S(\psi(t)),\qquad \tilde S=S(\Psi^{-1}(\psi(t))
\]
with the new $2\pi$-periodic waveform $\tilde S(\psi)=\tilde S(\psi+2\pi)$. 
Generally, there is no particular reason to prefer the representation
in terms of the waveform $S$ and the phase $\varphi$ to the representation in terms of the waveform $\tilde S$ and of the phase $\psi$. 

In applications, ``preferable'' phase variables can exist. For example, for a dynamical system
where the equation for the phase in form \eqref{eq:dynph} is used, the ``true'' phase has the property
of a uniform (on average) rotation, so that the distribution density of this phase does not depend on the phase itself. See more discussion in Refs.~\cite{kralemann2008phase}, where general possible phase variables have been named ``protophases'', and
a transformation of form \eqref{eq:trp} from a protophase to the phase having uniform distribution
density  has been suggested as a
step
in the phase reconstruction. However, in other applications other conditions on the phase may be imposed. 
On the other hand, there can be a ``preferable'' waveform. For example, a waveform with one maximum and one minimum 
over the period, can be transformed to a cosine waveform.
The ambiguity of the phase definition results also in the ambiguity of the instantaneous frequency, if the latter is defined as the time derivative of the phase. 

Summarizing, we state that the decomposition  of a signal \eqref{eq:pmsig} is not unique, and in different methods,
described below, different protophases and correspondingly different waveforms will be obtained.  
We will, however, consider demodulation as successful, if at least one particular protophase and the 
corresponding waveform can be found. The success of demosulation means that a protophase $\psi(t)$
is found, such that values of $|x(\psi)-x(\psi+2\pi)|$ are very small. Quantitatively, we define the squared integrated error of demodulation in
Eq.~\eqref{eq:errdef} below.

\subsection{Link function approach}
\label{app:link}
There is another formulation of the demodulation problem, based on the \textit{link function} approach 
[M. Holschneider, private communication]. Given a modulated signal~\eqref{eq:pmsig}, one looks for
a function of time $L(t)$, called link function, such that
\[
x(t+L(t))=x(t)\;.
\]
If a protophase $\psi(t)$ of the signal is known, then one can define the link function as
$\psi(t+L(t))=\psi(t)+2\pi$, what can be resolved via the inverse function $t(\psi)$ as $L(t)=t(\psi(t)+2\pi)-t$.
One can easily see that the link function is independent on the choice of the protophase, and in this sense 
is invariant. For the dynamical systems described in section~\ref{sec:pmdo}, the link function gives the 
return time for the Poincar\'e map from the value of the phase to the next value shifted by $2\pi$. The task of finding a protophase if the link function is known, is, to the best of our knowledge, not solved.

\section{Proxi-phase extraction} \label{sec:ppe}
\label{sec:phase_extraction}
Our main method is based on an iteration procedure. At each iteration stage, an 
approximation to the phase is extracted from the phase-modulated 
signal $x(t)$; we call the result of this extraction ``proxi-phase'' below. In fact, this extraction step 
is performed only once in most existing approaches to extract the phase dynamics from data (for example, see \citep{benitez2001use, blaha2011reconstruction, kralemann2008phase, topccu2018disentangling}).

The usual approach is to perform a two-dimensional embedding $x(t)\to (x(t),y(t))$, so that a 
phase-modulated signal is 
represented by ``rotations'' or ``loops'' on the plane $(x,y)$. In the field of 
dynamical system reconstruction, different
choices for $y$ 
have been discussed, two popular are a delayed signal $y=x(t-\tau)$ and the derivative 
$y=\frac{dx}{dt}$~\cite{Kantz-Schreiber-04,kim1999nonlinear,letellier1998non}. However,
the mostly stable and reliable results are achieved with the Hilbert transform~\cite{rosenblum2001phase}:
\begin{equation}
y(t)=\hat{H}[x(t)]=\frac{1}{\pi}P.V.\int_{-\infty}^\infty \frac{x(t')}{t-t'}dt'\;.
\label{eq:htdef}
\end{equation}
The resulting trajectory on the plane $(x(t),y(t))$ we call HT-embedding.

Below we will consider two basic ways to extract a proxi-phase from the HT embedding.
In the simplest case, where the waveform is close to a $\cos(\varphi)$ function, the HT embedding is close to a circle,
surrounding the origin (see Fig.~\ref{fig:embpl}(a) below),
and one can define the proxi-phase $\theta^{(a)}(t)$ as the angle from the origin (or another point close to the origin)
to the position of the signal:
\begin{equation}
\theta^{(a)}(t)=\text{arg}[x(t)+iy(t)]\;.
\label{eq:proxiphatan}
\end{equation}
This definition corresponds to the proxi-phase as the argument of the analytic signal $z(t)=x(t)+iy(t)$.
Therefore we will call $\theta^{(a)}$ the \textit{analytic proxi-phase}. The proxi-phase $\theta^{(a)}$
should be additionally unwrapped, so that it is a motonously growing function of time.

This simple definition of the proxi-phase will not work
for a complex waveform with several maxima and minima on the period. 
In this case the embedding on the plane $(x,y)$ has many loops (see Fig.~\ref{fig:embpl}(e) below), 
and the evaluation
of the proxi-phase according to \eqref{eq:proxiphatan} is not possible. 
Even for the simple waveform $S(\varphi)=\cos(\varphi)$, if the phase modulation is strong and the
modulation frequency is high,  a situation may occur where the analytic proxi-phase according to~\eqref{eq:proxiphatan} is non-monotonous. This makes application of the analytic proxi-phase impossible for such cases.

A more universal proxi-phase definition was suggested in 
Ref.~\cite{kralemann2008phase} and adopted for the analysis of an 
electrocardiogram signal in Ref.~\cite{Kralemann_etal-13}.
This proxi-phase is based on the curve length, to avoid the mentioned problems. 
The length-based definition of a proxi-phase $\theta^{(b)}(t)$ is
\begin{equation}
\theta^{(b)} (T) = \int_0^T\; dt\; \sqrt{\left(\frac{d x}{dt}\right)^2+\left(\frac{d y}{dt}\right)^2}\;.
\label{eq:linear}
\end{equation}
For any sufficiently smooth signal, this definition provides a monotonously growing \textit{length proxi-phase},
even if the embedded trajectory has loops.

A small drawback of this definition is that the proxi-phase $\theta^{(b)}(t)$ is not normalized to intervals of $2\pi$ at each rotation,
contrary to the analytic proxi-phase~\eqref{eq:proxiphatan}. 
This does not affect further iteration steps to be described below, 
because the HT is invariant under rescaling of the function argument. At the very end
of the procedure, one can perform the normalization 
by multiplying $\theta^{(b)}$ with a factor $2\pi \hat{L}^{-1}$, if needed. Here $\hat{L}$ is the average period 
of the signal in terms of the length~\eqref{eq:linear} after the iteration procedure. 
To define it, we first define the normalized periodicity error of a function
$x(\theta)$, with respect to a period guess $L$, as
\begin{equation}
\tilde\varepsilon(L) = \frac{\int_{\theta_{min}}^{\theta_{max}} (x(\theta'+L) - x(\theta'))^2 d\theta'}
{\int_{\theta_{min}}^{\theta_{max}} 
x^2(\theta') d\theta'}\;.
\label{eq:errdef}
\end{equation}
In case of the analytic proxi-phase, we can set $L=2\pi$, then expression~\eqref{eq:errdef} provides
the quality of the phase reconstruction. $\varepsilon=\tilde\varepsilon(2\pi)$ 
only weakly dependent on the protophase definition. 
If it tends to zero for one protophase, then it tends
to zero for other protophases as well.

For the length proxi-phase,
the value of $L$ is not known a priori. Here, we obtain an ``optimal'' 
period by minimizing the error \eqref{eq:errdef} with respect to $L$:
\[
\hat{L}=\text{argmin\ }\tilde\varepsilon(L)\;,\qquad \varepsilon=\min_{L}\tilde\varepsilon(L)=\tilde\varepsilon(\hat{L})\;.
\] 
This error will be  used in figures \ref{fig:pererr1} and \ref{fig:pererr2} below to quantify quality of demodulation. 

The proxi-phase extraction can be viewed as a (nonlinear) operator $\hat{P}$, providing for a given
signal $x(t)$ a proxi-phase 
\[
\theta(t)=\hat{P}[x(t)]
\]
with one of the methods above.

\section{Iterated Hilbert transform embeddings}
\label{sec:ihe}

Application of the proxi-phase extraction by one of the 
methods described in section~\ref{sec:phase_extraction}
above shows that this proxi-phase is not the phase, because in the function  $x(\theta)$ some phase modulation is still present such that it is not exactly periodic.

We suggest to proceed in an iterative manner: We consider
the obtained proxi-phase (which we supply with an index that counts the steps in the iteration process)
$\theta_1$ as a new time variable, and apply the proxi-phase extraction to the signal $x(\theta_1)$. Namely,
we calculate $\theta_2 (\theta_1)=  \hat{P}[x(\theta_1)]$, and so on. In other words,
we perform iterations
\begin{equation}
\theta_{n+1} (\theta_{n})=  \hat{P}[x(\theta_{n})]
\label{eq:iter1}
\end{equation}
The original time can be treated as the zero-index proxi-phase $\theta_0=t$.
To illustrate, we write explicitly the steps for the analytic signal proxi-phase $\theta^{(a)}$:
\begin{equation}
\begin{gathered}
y(\theta_{n})=\hat{H}[x(\theta_n)]=\frac{1}{\pi}\text{PV}\int_{-\infty}^\infty\frac{x(\theta_n')}{\theta_n-\theta_n'}
d\theta_n',\\ 
\theta_{n+1}=\text{arg } [x(\theta_n)+iy(\theta_n)]\;.
\end{gathered}
\label{eq:iter2}
\end{equation}
In numerical implementation of the iterations, one has to perform the HT for a function defined
on a non-uniform grid; see~\ref{ap:nonun} for the implementation we used.

\begin{figure}[!tbp]
\centering
\includegraphics[width=1.0\columnwidth]{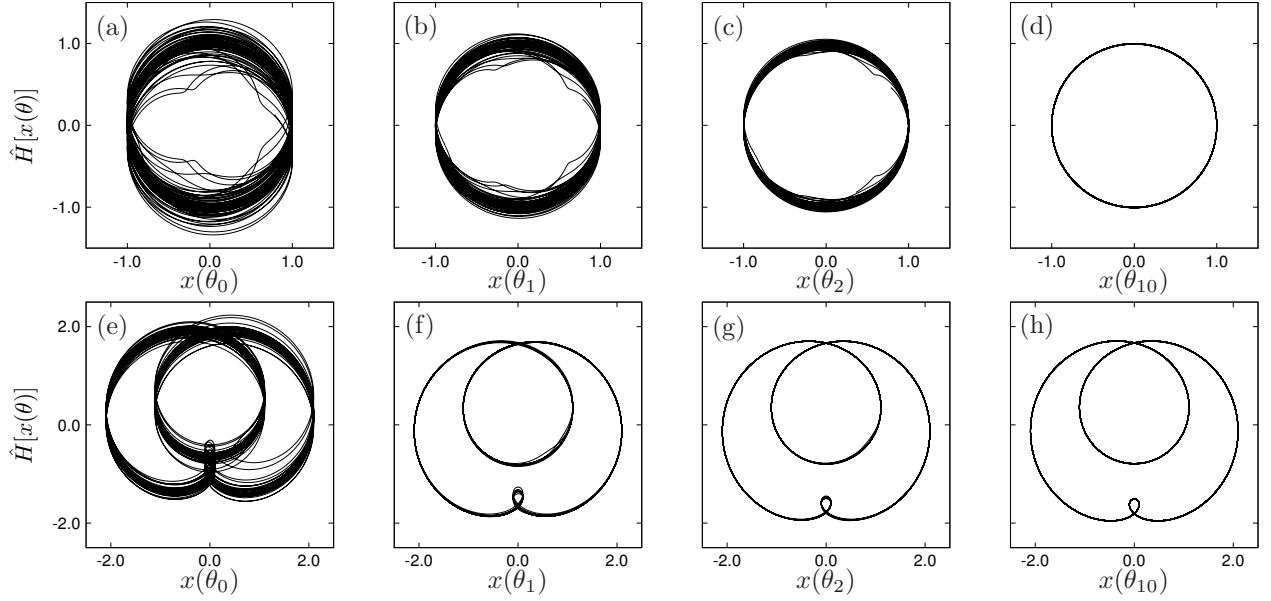}\\
\caption{
Iterated HT embeddings for the phase-modulated signals $x_{1,2}(t)=S_{1,2}(\varphi(t))$
(see expressions~(\ref{eq:swf},\ref{eq:cwf})) with modulation $\varphi(t)=t+1.2(\sin(0.25\sqrt{2}t) + \cos(0.25\sqrt{3}t))$. Panels (a)-(d) show
the simple waveform case, here the analytic proxi-phase was adopted. Panels (e)-(h) show the complex waveform, here the length proxi-phase was used. The iteration steps: (a),(e): $n=0$; (b),(f): $n=1$;
(c),(g): $n=2$, and (d),(h): $n=10$.
}
\label{fig:embpl}
\end{figure}

Our numerical simulations show that this iterative 
procedure effectively reduces the errors \eqref{eq:errdef}, and, 
with high accuracy, $\theta_n$ converges to a correct protophase 
for large $n$. Just visually one can check this
by comparison of embedded plots $(x(\theta_n),y(\theta_n))$ for 
different iteration steps $n$, see Fig.~\ref{fig:embpl}. If the signal
$x(\theta_n)$ is a periodic function of $\theta_n$, then $y(\theta_n)$ is also a periodic function, and
on the embedding plane one observes just a closed curve. Contrary to this, if $\theta_n$ is just an 
approximation and $x(\theta_n)$ is not periodic but contains rest modulation, then one
observes a non-closed trajectory forming a sequence of non-equal loops.

In Fig.~\ref{fig:embpl} one can see that just one step in the phase extraction does not
provide a good phase -- the embeddings in panels (b),(f) look like quite wide bands. In course of iterations,
these  bands become narrower, and already at the $10$-th step one observes very narrow lines (panels (d),(h)),
what means that the demodulation procedure is successful.

Below we first present a theory, showing that the proposed iteration 
method indeed converges for a cosine waveform and a weak modulation.  
Then, in section~\ref{sec:numtests}, we will first check predictions of the theory; and then explore
convergence of the method for strong modulations, by virtue of the error analysis according
to expression~\eqref{eq:errdef}.

\section{Theory}\label{sec:theory}

Our strategy here is to show that at each iteration~\eqref{eq:iter1} the residue 
between the true phase and the calculated phase decreases. We represent 
the modulation of the true phase as
\[
\varphi(t)=t+\epsilon q(t),\qquad \epsilon\ll 1\;,
\]
and restrict our analysis to terms of first order 
in $\epsilon$. We have fixed the basic frequency of the signal to one, without loss of generality. 
Thus, the signal and its HT are represented as 
\begin{equation}
\begin{aligned}
x(t)  = & S(\varphi(t)) \approx S(t) + \epsilon S'(t)q(t) \;,\\
\hat H[x](t)  \approx & \hat H[S(t)] + \epsilon \hat H[S'(t)q(t)]\;,
\end{aligned}
\label{eq: appprox sig and Hilbert}
\end{equation}
where prime denotes derivative.
We calculate  both proxi-phases in the first order in $\epsilon$ based on~\eqref{eq:proxiphatan} and~\eqref{eq:linear} 
respectively, and obtain general expressions, valid for arbitrary waveforms $S$ and modulations $q(t)$: 
\begin{equation}
\begin{aligned}
\theta^{(a)}(t) & =  \arctan\left( \frac{\hat H[S(t)]}{S(t)} \right) + 
\epsilon\frac{\hat{H}[S'(t)q(t)]S(t)-S'(t)q(t)\hat{H}[S(t)]}{(S(t))^2+H[S(t)]^2}\\
\theta^{(b)}(t) & = \int_0^t \sqrt{(S'(\tau))^2+\hat H[S'(\tau)]^2} d\tau\\&
+ \epsilon \int_0^t \frac{S'(\tau)(q'(\tau)S'(\tau) + S''(\tau)q(\tau)) + 
\hat H[S'(\tau)](\hat H[q'(\tau)S'(\tau)] + \hat H[q(\tau)S''(\tau)])}{\sqrt{(S(\tau))^2
+\hat H[S(\tau)]^2}} d\tau 
\label{eq:twoph}
\end{aligned}
\end{equation}
We are able to give compact tractable expressions for the simplest 
waveform $S(\varphi)=\cos\varphi$ only. This case is treated below.

The difference $\Delta^{(a,b)}(t)=\varphi(t)-\theta^{(a,b)}(t)$ between the true phase and the reconstructed one 
should be interpreted
as the rest modulation after 
the first iteration. For the analytic proxi-phase~\eqref{eq:proxiphatan} and the length 
based proxi-phase~\eqref{eq:linear}, in the case $S(\varphi)=\cos(\varphi)$, the expressions 
for the rest modulation, obtained through simple calculations from \eqref{eq:twoph}, read
\begin{gather}
\Delta^{(a)}(t)  =  \epsilon \left(q(t)+\cos(t)\hat{H}[q(t) \sin(t)] - q(t) \sin^2(t)\right) \label{eq:err analytic phase}\;, \\
\Delta^{(b) \prime}(t) =  \frac{\epsilon}{2}q'(t) + \frac{\epsilon}{2}[\cos(2t) q'(t) - \sin(2t) q(t)
  + 2\cos(t) \hat H[q'(t)\sin(t)] + 2\cos(t)\hat H[q(t)\cos(t)]]\;.
 \label{eq: err linear phase}
\end{gather}
Here, to get rid of the integral in the expression for $\Delta^{(b)}$, we take derivative with respect to time $t$.

We now use the Bedrosian's theorem~\cite{bedrosian1962product}, which states that the HT
 of a product of a low-frequency  $Q_L(t)$ and a high-frequency $Q_H(t)$ 
signals and can be represented as
\[
\hat{H}[Q_L(t) Q_H(t)]=Q_L(t)\hat{H}[Q_H(t)]\;.
\]
The condition for the Bedrosian's theorem is that the spectra of $Q_L$ and $Q_H$ do not overlap:
the whole spectrum of $Q_L$ lies left to the spectrum of $Q_H$.

Thus, to evaluate the terms 
$\hat{H}[q(t) \sin(t)]$, $\hat{H}[q(t) \cos(t)]$, and $\hat{H}[q'(t) \sin(t)]$, we represent $q(t)$ as a sum of two components
$q(t)=\ell(t)+h(t)$, where the Fourier spectrum 
of the slow component $\ell(t)$ contains
frequencies $|\omega| <1$, while the Fourier spectrum 
of the fast component $h(t)$
contains frequencies that obey $|\omega|>1$. Then, we obtain
\[
\hat{H}[q(t) \sin t]=\hat{H}[(\ell(t)+h(t)) \sin(t)]=-\ell(t)\cos(t)+\hat{H}[h(t)]\sin(t) \;,
\]
and similar for $\hat{H}[q(t) \cos(t)]$ and $\hat{H}[q'(t) \sin(t)]$. Substituting this
in~\eqref{eq:err analytic phase},\eqref{eq: err linear phase}, we see that the component 
$\ell(t)$
does not contribute to the rest modulation. Only the contribution of the high-frequency component remains:
\begin{gather}
\Delta^{(a)}(t)  =  \frac{\epsilon}{2}h(t)+\frac{\epsilon}{2}
\left(h(t) \cos (2 t) +\sin (2 t) \hat{H}[h(t)]\right)\;, \label{eq: bedrosian err analytic}\\
\Delta^{(b)\prime}(t)  = \frac{\epsilon}{2} h'(t) + \frac{\epsilon}{2}\left(\hat H[h(t)] + 
\cos(2t) h'(t) - \sin(2t) h(t) 
  + \sin(2t) \hat H[h'(t)] + \cos(2t) \hat H[h(t)]\right)\;.
\label{eq: bedrosian error length}
\end{gather}

At this stage, it is more convenient to operate with the Fourier spectra 
$\mathcal{F}_h(\omega),\mathcal{F}_\Delta(\omega)$ of functions $\epsilon h(t),\Delta(t)$.
The Hibert transform in the Fourier space has the form $\mathcal{F}(\hat{H}[h(t)])(\omega)=-i\text{\ sgn}(\omega)
\mathcal{F}_h(\omega)$. Application of the Fourier transform
to expressions \eqref{eq: bedrosian err analytic} and \eqref{eq: bedrosian error length} yields:
\begin{equation}
\begin{aligned}
\mathcal{F}_{\Delta^{(a)}}(\omega)=&\frac{1}{2}
\left[\mathcal{F}_h(\omega)+\frac{1}{2}(1-\text{sgn}(\omega-2))
\mathcal{F}_h(\omega-2)+\frac{1}{2}(1+\text{sgn}(\omega+2))
\mathcal{F}_h(\omega+2)\right]\;,\\
\mathcal{F}_{\Delta^{(b)}}(\omega)=&\frac{1}{2}
\left[\frac{\omega-\text{sgn}(\omega)}{\omega}\mathcal{F}_h(\omega)+
\left(\frac{\omega-1}{\omega}(1-\text{sgn}(\omega-2))\mathcal{F}_h(\omega-2)\right.\right.+\\&
\left.\left.\frac{\omega+1}{\omega}(1+\text{sgn}(\omega+2))
\mathcal{F}_h(\omega+2)\right)\right]\;.
\end{aligned}
\label{eq:ftr}
\end{equation}
We now use that $\mathcal{F}_h(\omega)=0$ for $|\omega|<1$, what allows us to simplify expressions~\eqref{eq:ftr}
to
\begin{equation}
\mathcal{F}_{\Delta}(\omega)=\begin{cases}
A(\omega)\mathcal{F}_h(\omega)+B(\omega)\mathcal{F}_h(\omega+2)&\text{if }\omega>1\;,\\
B(\omega)\mathcal{F}_h(\omega+2)+A(\omega)\mathcal{F}_h(\omega-2)&\text{if }-1<\omega<1\;,\\
B(\omega)\mathcal{F}_h(\omega)+A(\omega)\mathcal{F}_h(\omega-2)&\text{if }\omega<-1\;,
\end{cases} \label{eq: fourier analytic phase}
\end{equation} where 
\begin{equation}
\begin{aligned}
A(\omega)=B(\omega)=1/2 &
\text{ for the analytic proxi-phase }\Delta^{(a)},\\
A(\omega)=\frac{\omega \mp 1}{2\omega},\;\; B(\omega)=\frac{\omega \pm 1}{2\omega} &
\text{ for  the length proxi-phase }\Delta^{(b)}.
\end{aligned}
\label{eq:damp}
\end{equation}
In~\eqref{eq:damp}, the upper sign is used for $\omega>1$ and the lower sign is used for $\omega<-1$.

\begin{figure}[h!tbp!]
\centering
\includegraphics[width=0.5\columnwidth]{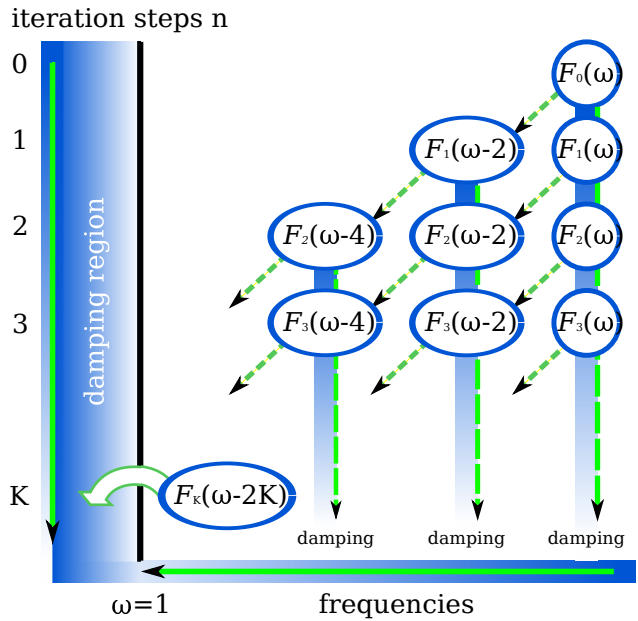}
\caption{Depicted is the damping of Fourier modes in the range 
$2K<\omega<2K+2$ for the simple waveform $S(\varphi) = \cos(\varphi)$. 
In each iteration step, the method damps $\mathcal{F}_n(\omega)$ 
(dashed vertical arrows) and additionally generates 
a new Fourier mode at a smaller frequency (dashed diagonal arrows). At step $K$, a Fourier mode
with frequency less than 1  
is generated in the low-frequency region (bold green-white arrow), 
where it disappears before the next iteration. 
The damping factors are given by~\eqref{eq:damp}.}
\label{fig:hevol}
\end{figure}

Relation~\eqref{eq: fourier analytic phase} can be considered as a transformation of the rest phase modulation
at a step of our iteration procedure. Denoting the spectrum of the high-frequency modulation 
$\epsilon h(t)$ at the $n$-th step
of the iteration procedure as $\mathcal{F}_{n}$, we can write a general recursion
formula for the evolution of this spectrum under
iterations:
\begin{equation}
\mathcal{F}_{n+1}(\omega)=\begin{cases}
A(\omega)\mathcal{F}_{n}(\omega)+B(\omega)\mathcal{F}_{n}(\omega+2)&\text{if }\omega>1,\\
B(\omega)\mathcal{F}_{n}(\omega+2) + A(\omega)\mathcal{F}_{n}(\omega-2)&\text{if }-1<\omega<1,\\
B(\omega)\mathcal{F}_{n}(\omega) + A(\omega)\mathcal{F}_{n}(\omega-2)
&\text{if }\omega<-1.
\end{cases}
\label{eq:fiter}
\end{equation}
This relation is the main result of our weak modulation analysis. We now discuss its meaning for the iteration procedure of demodulation (where we focus, due to the evident symmetry, on positive
frequencies only).
\begin{enumerate}
\item The first observation is that in both cases of proxi-phase definition, slow modulation (i.e. with frequencies
lower than the basic oscillation frequency) is resolved exactly already in the first iteration. This is
a well-known empirical fact that the traditional HT phase extraction (where only the first iteration is used)
works well for slow modulation, but is not so good for fast modulation.
\item The high-frequency components do not disappear immediately, but are, on one hand, damped by 
factor $A(\omega)$, and 
on the other hand, produce components shifted to lower frequencies and damped by 
factor $B(\omega)$. Schematically, this 
is illustrated in Fig.~\ref{fig:hevol}. This scheme shows that for all spectra decaying at infinity, i.e. with
$\lim_{\omega\to\infty} |\mathcal{F}_{0}(\omega)|=0$, the modulation eventually disappears, i.e. 
$\lim_{n\to\infty}|\mathcal{F}_{n}(\omega)|=0$. For some
important classes of spectra, one can show that the eigenvalues of the transformation operator~\eqref{eq:fiter}
are smaller than one (see ~\ref{app:eval}), so that the iteration procedure converges exponentially.
In particular, for spectra $\mathcal{F}_h(\w)$ which decay exponentially $\mathcal{F}_h(\w)\sim 
\exp[-\alpha \omega]$ (what corresponds to a smooth modulation $h(t)$), we 
have $|\mathcal{F}_n|\sim \left(\nicefrac{1+\exp[-2\alpha]}{2}\right)^n\underset{n\to\infty}{\to} 0$.
\item In the analysis above we have not considered components of modulation having frequencies
$1,3,5,...$, i.e. harmonics of the basic oscillation frequency. Such components cannot be demodulated, because 
one cannot distinguish them from the original phase -- in fact, a modulation with these frequencies is
equivalent to changing the shape of the waveform $S(\varphi)$.
\end{enumerate}

\section{Numerical tests}
\label{sec:numtests}

\subsection{Testing theoretical relations} \label{sec: spectra}

\begin{figure}[!tbp!]
\centering
\includegraphics[width=\columnwidth]{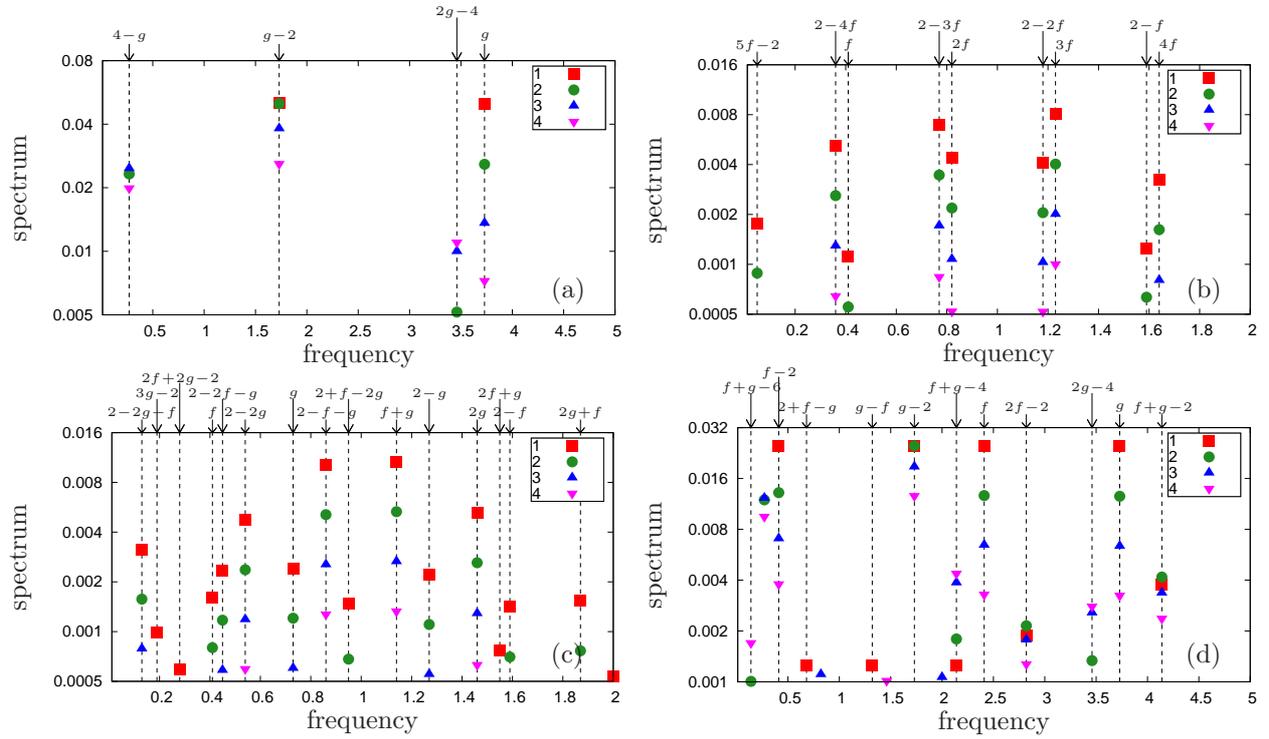}
\caption{We show the most essential spectral components of the phase demodulation
error $\theta^{(a)}_n-\varphi$ for the first 4 iterations 
(squares, circles, up and bottom triangles, respectively). The observed
spectral components are marked with arrows on top of panels.
In all cases the 
original signal is $x(t)=\cos(\varphi(t))$, where the modulation is 
given by Eq.~\eqref{eq:tfm}. The 
vertical scale is logarithmic, with the tics marking factor 2, to allow for a visual
comparison with the theory, where the damping factors are $1/2$.}
\label{fig:err-fourier}
\end{figure}

Our first numerical example is intended to illustrate the theory of Section~\ref{sec:theory}. 
Therefore, we use here the simple waveform $S(\varphi)=\cos (\varphi)$, 
and the modulation is a function of time
\begin{equation}
\varphi(t)=t+a\sin(f t)+b \cos(g t)\;,
\label{eq:tfm}
\end{equation}
possesing two frequencies $f$ and $g$. We performed iterative HT embeddings
as described above, calculated at each step the analytic proxi-phase, and found 
the spectral components of the differences $\theta^{(a)}_n(t)-\varphi(t)$. 
These spectral components are presented in Fig.~\ref{fig:err-fourier} for four numerical setups. 

\begin{itemize}
\item[Case (a).] Panel (a) of Fig.~\ref{fig:err-fourier} shows the case $a=0$, $b=0.2$, $g=3.73$
of a single-harmonic modulation with high frequency. Here 
the modulation is relatively weak, and we expect relations~\eqref{eq: fourier analytic phase} to hold. One 
can indeed see that, in agreement with the theory,  
$\theta^{(a)}_1-\varphi$ contains harmonics at frequencies $g$ and $g-2$
with the same amplitude. At the next iteration, the component at $g$ decreases by factor 2,
while the component at $g-2$ remains the same; additionally a component with frequency
$|g-4|$ appears. Also at iterations 3 and 4, relation \eqref{eq:fiter} is well fulfilled.
\item[Case (b).] Panel (b) illustrates nonlinear effects at a strong low-frequency
single-harmonic modulation.
Here $a=1$, $f=0.41$, and $b=0$. One can see that although the frequency of modulation is 
less than one, this component at $f$ does not disappear completely in $\theta^{(a)}_1-\varphi$
as predicted by the linear theory, but is finite (already rather small). In the first iteration
higher harmonics $nf$ and the combinational frequencies $nf-2$ appear, and they are 
relatively large. During next iterations the amplitudes of these components decrease with 
a factor $\approx 1/2$.
\item[Case (c).] Panel (c) shows the case of relatively strong two-frequency modulation, 
where both frequencies are low: $a=b=0.3$, $f=0.41$, $g=0.73$. Due to nonlinearity, many combinational frequencies $kf+lg$, $kl+fg-2$ with integers
$k$ and $l$ are exited. It appears that all the new frequencies, like in case (b), are created in the first iteration, 
further iterations follow roughly the law \eqref{eq:fiter}.
\item[Case (d).] Panel (d) shows, like case (c), a strong two-frequency modulation, but with
high basic frequencies $a=b=0.1$, $f=2.41$, $g=3.73$. One can see that at some combinational 
frequencies, the level even initially increases due to the cascade process of amplitude shifting 
(see, e.g., components with $f+g-4$), and one needs more iterations to reduce error at these components. 
\end{itemize}

In conclusion, the presented numerical tests show that the theory based on the linear 
approximation
works well for weak modulation, but for strong modulation, there are essential nonlinear 
effects at the first few iterations. Because the effective modulations become 
weaker in the course of iterations, the theory better describes higher iterations.

\subsection{Accuracy of the iterative procedure}\label{sec: period errors}

In the following, we present numerical results for the signals $x(t)=S_{1,2}(\varphi(t))$,
where the waveforms are given by expressions~\eqref{eq:swf},\eqref{eq:cwf}, and
the phase modulation is a quasiperiodic function of time
\begin{equation}
\varphi(t) = t + \frac{a}{R} \cos (R \omega_1 t) + \frac{b}{R} \cos(R\omega_2 t) \;.
\label{eq:phmf}
\end{equation}
We have chosen $\omega_1=\sqrt{2},\;\omega_2=\sqrt{3},\;a=b=0.3$. Parameter $R$ allows
us to vary the basic frequencies, keeping the range of the phase modulation 
(which is quite large,  $\text{min}(\dot \varphi) = 1 - a(\omega_1+\omega_2) \approx 0.056$,
so the first-order theoretical analysis of section~\ref{sec:theory} is not applicable) 
constant. In all runs we used
discretisation step $\Delta t=0.002$, the length of the time series was about 95 basic periods. 
Convergence was characterized by evaluating error~\eqref{eq:errdef}, where, to discard boundar effects, only central 80\% of the time series where used.

Our goal is to quantify performance of the iterative procedure described, in dependence 
on the characteristic time scale defined by parameter $R$, on the properties of the
waveform (simple vs complex), and on the choice of proxi-phase in the iterative process 
(analytic vs length-based).


\subsubsection{Simple waveform: comparison of analytic and length-based proxi-phases}

\begin{figure}[h!tbp!]
\centering
\includegraphics[width=0.6\columnwidth]{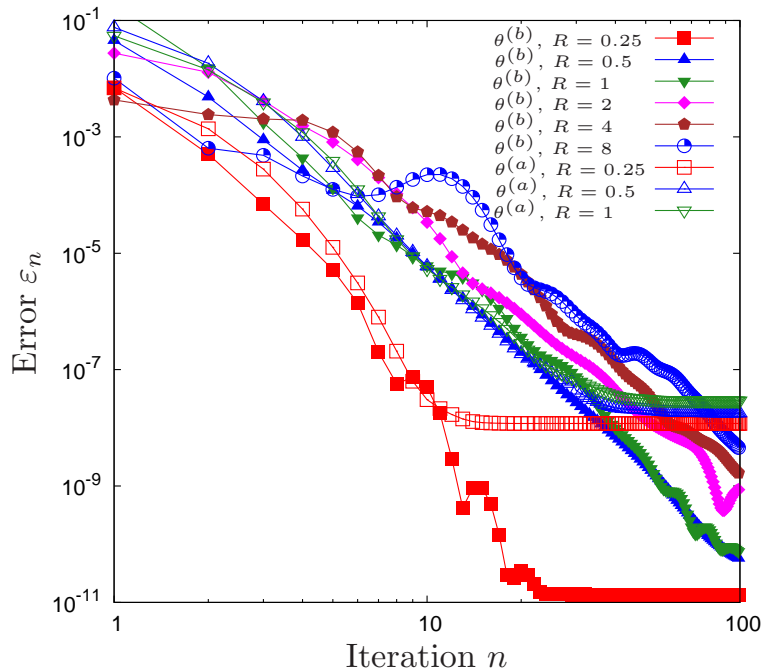}
\caption{Errors of demodulation for simple waveform $S_1$ and both proxi-phases 
calculations in dependence on the iteration index $n$ and $R$. For $R=0.25,\;0.5,\;1$,
results for the analytic and length proxiphases are depicted in the same color, but with 
open and filled markers, respectively.} 
\label{fig:pererr1}
\end{figure}

For the simple
 waveform~\eqref{eq:swf}, one can use both proxi-phases - the analytic one
and the length-based one. The results are shown in Fig.~\ref{fig:pererr1}.
One can see that both proxi-phases 
ensure demodulation with a very small rest error, if the modulation is slow ($R\leq 1$). 
For the lowest-frequency case, the length method 
performs even significantly better at high iterations.

For larger values of $R\geq 2$, the length method appears to be the only applicable one. 
Here, the analytic proxi-phase fails because
at one of the first iterations
the analytic proxi-phase becomes a non-monotonous function of time, 
and iterations break. In contradistinction, the length-based 
phase always provides a monotonous estimate of the phase. 
One can see that for the fastest modulation explored ($R=8$), the error decreases
not monotonously with $n$ (it increases at $n\gtrsim 10$), but even in
this case, the error of the iteration procedure eventually becomes rather small.

\subsubsection{Complex waveform}
\label{sec:cwf}

\begin{figure}[!tbp!]
\centering
\includegraphics[width=0.6\columnwidth]{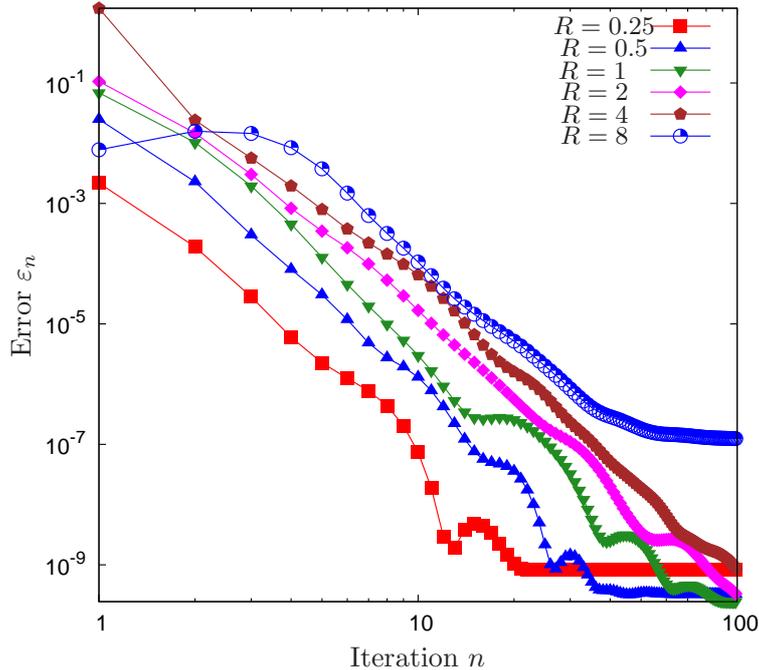}
\caption{Errors of demodulation in dependence on the iteration index $n$ and $R$
for the complex waveform $S_2$.}
\label{fig:pererr2}
\end{figure}

For the complex waveform~\eqref{eq:cwf}, only the length-based phase can be used. 
In Fig.~\ref{fig:pererr2} we show the errors vs the iteration number $n$, for
the same values of the parameter $R$ as in Fig.~\ref{fig:pererr1}. 
In all cases the 
final error is very small, close to $10^{-7}-10^{-8}$. Similarly to the 
case of the simple waveform, here at the fastest example ($R=8$), the demodulation 
is not monotonous. 

The presented results confirm that the 
described method provides 
an effective demodulation of the complex signal. We attribute the rest error to 
numerical inaccuracies in calculation of the HT, of the length of the 
trajectory on the plane $(x,y)$, 
and in the evaluation of the error itself. This is supported by the observation that
the asymptotic errors grow significantly if a larger discretization step is used.

It is important to mention, that contrary to 
the analytic proxi-phase~\eqref{eq:proxiphatan} which is $2\pi$-periodic, the 
length-based proxi-phase \eqref{eq:linear} does not have a ``natural'' period: 
strictly speaking, a proper normalization to period $2\pi$ is only possible 
at the end stage of the iteration procedure, where the 
periodicity of $x(\theta^{(b)})$ is well-defined.

\subsection{Arbitrary phase modulation} \label{sec: wild phase}

\begin{figure}[!tbp!]
\centering
\includegraphics[width=\columnwidth]{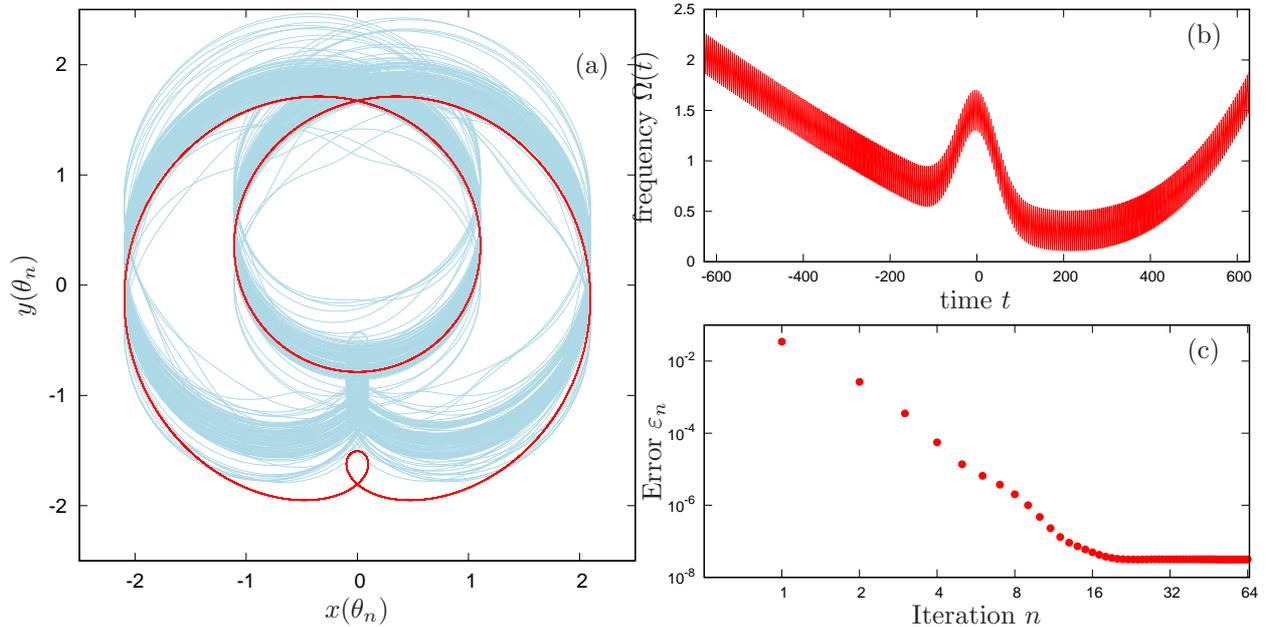}
\caption{Demodulation of a signal having 
complex waveform $S_2$  and the instantaneous frequency \eqref{eq:omega} (panel (b)). 
Panel (a) shows embeddings for the original signal
$x(\theta_0)$ (blue line) and for the iteration $x(\theta_{64})$ (red line). 
Panel (c): Errors in dependence on the iteration number.}
\label{fig:err_wph}
\end{figure}

Above we considered phase modulations as quasiperiodic deviations from the linear phase 
growth. The method is, however, not restricted to such cases. As an example, we present here results for the complex waveform $S_2$ (Eq.~\eqref{eq:cwf}) with 
phase modulation defined as $\dot\varphi=\Omega(t)$,
with a rather arbitrarily chosen instantaneous frequency
\begin{equation}
\Omega(t) = -2 \tau + 0.5\exp(2 \tau) + \exp[-100 (\tau)^2] + 0.2\cos(\sqrt{2}600  \tau)\;,
 \qquad \tau = \frac{t}{200\pi} \;,\quad -200\pi\leq t<200\pi\;,
\label{eq:omega}
\end{equation}
see Fig.~\ref{fig:err_wph}(b).
For the sampling with $\Delta t=0.005$, the demodulation 
error $\approx 3\cdot 10^{-8}$ was achieved after 20 iterations (see Fig.~\ref{fig:err_wph}(c)). 
In Fig.~\ref{fig:err_wph}(a) the quality of demodulation is illustrated visually, like in 
Fig.~\ref{fig:embpl}.

\section{Transformation of a protophase to true phase}
\label{sec:prot}

\begin{figure}[h!tbp!]
\centering
 \includegraphics[width=0.4\columnwidth]{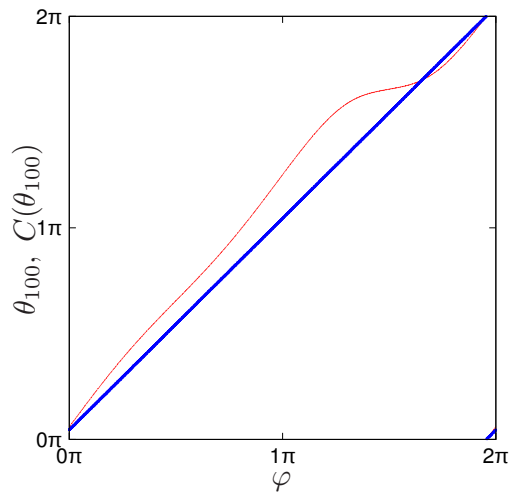}
\caption{Red points: values of the reconstructed protophase $\theta^{(b)}_{100}$
at $100$-th iteration vs. the true phase $\varphi$. 
Blue bold points: the transformed protophase $C(\theta_{100})$. One can see that 
the relation between $C(\theta_{100})$ and $\varphi$ is nearly linear. }
\label{fig:clph}
\end{figure}

As discussed in section~\ref{sec:ppvp} above, there is an intrinsic ambiguity in the
phase demodulation problem: one can perform a transformation of a waveform together
with a transformation of an obtained protophase, such that the signal remains the same.
Thus, one needs additional conditions to specify a particular phase and a particular 
waveform uniquely.
In the context of the dynamical system applications (section~\ref{sec:pmdo}), a natural condition
is that the phase should grow on average uniformly in time. This condition was
adopted in \cite{kralemann2008phase}, where a procedure for the transformation
$\theta\to C(\theta)$ from
a protophase to the uniformly on average growing phase was described. Roughly
speaking, for a wrapped
protophase one determines the probability distribution density on the interval $[0,2\pi)$,
and performes a transformation such that the new phase has the uniform 
distribution density. To illustrate this approach, we take an example from section~\ref{sec:cwf} 
with $x(t)=S_2(\varphi(t))$, where $S_2$ is given by Eq.~\eqref{eq:cwf} and
$\varphi(t)$ is given by Eq.~\eqref{eq:tfm} with $R=8$.
In Fig.~\ref{fig:clph} we show the protophase obtained via the length proxi-phase method, and its transformation
according to method~\cite{kralemann2008phase}. The transformed phase is close to the original
one, up to an unavoidable, but unimportant, shift. We stress here, that the accuracy of the transformation
$\theta\to C(\theta)$ is rather weak, because it relies on the empirical probability density, 
which one obtains based on  finite-size data. 

\section{Conclusion}
\label{sec:concl}

In summary, we have proposed a method of phase demodulation based on the iterative HT
embeddings.
In all tested cases it leads to an effective accurate reconstruction of the waveform and of the phase of
the signal, up to unavoidable ambiguity due to protophase variety. Our method can be applied
to simple and complex waveforms. In all cases we have found that the proxi-phase extraction
based on the length of the embedded trajectory is superior to a traditional definition 
based on the analytic signal.   

We have performed theoretical analysis valid for a weak modulation of
the simplest cosine waveform. This analysis shows that while slow modulation is
detected exactly already within the first iteration, for fast components the iterations are
indeed needed. Furthermore, this analysis yields exponential (in number of iterations) 
convergence rate for  modulation signals which 
spectra decay exponentially at high frequencies. We have shown also
that nonlinear effects in the case of not so weak modulation lead to appearance of
combinational frequencies in the modulation spectrum, what nevertheless does not prevent
eventual convergence of the procedure.
 
In this study we considered only regular signals and used a quite high sampling rate.
This is mainly to achieve the highest possible performance, because larger discretization
steps lead to errors in calculations of the Hilbert transforms. The analysis of different
factors affecting the accuracy will be subject of further research. Also applications
of the method to practical situations of signals from forced dynamical systems are
now under consideration and will be reported elsewhere.

\section*{Acknowledgement}
A. P. was supported by Russian Science Foundation (Grant No. 17-12-01534).
E. G. was supported by Friedrich-Ebert Stiftung. 
We thank Michael Rosenblum, 
Michael Feldman, and Mathias Holschneider 
for their advices and fruitful discussions.

\section*{Bibliography}

\begin{appendix}

\section{Numerical implementation of the Hilbert transform on a non-uniform grid}
\label{ap:nonun}

The HT of a function $u(\phi)$ is defined as
\[
H(\psi)=\frac{1}{\pi}\text{PV}\int_{-\infty}^\infty \frac{u(\phi)}{\psi-\phi}d\phi\;.
\]
Suppose the function $u$ is given on a discrete set of points $\phi_1,\phi_2,\ldots,\phi_N$ as values $u_1,u_2,\ldots$.
We now want to find $H_i$ in the same set of points $\psi_1=\phi_1,\ldots$

We apply the trapezoidal rule, similar to the approach of~\cite{Zhou_etal-09} for a uniform grid, and 
write the integral as a sum of elementary integrals over domains $(\phi_{k},\phi_{k+1})$:
\[
H(\phi_i)=\frac{1}{\pi}\sum_{k=1}^{N-1}\int_{\phi_k}^{\phi_{k+1}} \frac{u(x)}{\phi_i-x}dx=\frac{1}{\pi}
\sum_{k=1}^{N-1} I_k\;.
\]
We approximate $u(x)$ on the interval $(\phi_{k},\phi_{k+1})$  by a linear function:
\begin{gather*}
u(x)=u_k+\frac{(u_{k+1}-u_k)}{\phi_{k+1}-\phi_k}(x-\phi_k)=
a_k+b_k x\;,
\end{gather*}
where
\[
a_k=\frac{u_k\phi_{k+1}-u_{k+1}\phi_k}{\phi_{k+1}-\phi_k}\;,\qquad b_k=\frac{u_{k+1}-u_k}{\phi_{k+1}-\phi_k}\;.
\]
Thus, the elementary integrals are
\begin{gather*}
I_k=\int_{\phi_k}^{\phi_{k+1}}\frac{a_k+b_k x}{\phi_i-x}dx
=u_k-u_{k+1}-
\frac{u_k\phi_{k+1}-u_{k+1}\phi_k+\phi_i(u_{k+1}-u_k)}{\phi_{k+1}-\phi_k}\ln\left|\frac{\phi_{k+1}-\phi_i}{\phi_k-\phi_i}\right|\;.
\end{gather*}
This expression is valid  for $k\neq i$, $k\neq i-1$.

To calculate the integrals around $\phi_i$, we represent $u=u_i+d_+ (x-\phi_i)$ on $\phi_i<x<\phi_{i+1}$
and $u=u_i+d_- (x-\phi_i)$ on $\phi_{i-1}<x<\phi_i$, where
\[
d_+=\frac{u_{i+1}-u_i}{\phi_{i+1}-\phi_i},\qquad d_-=\frac{u_i-u_{i-1}}{\phi_i-\phi_{i-1}}.
\]
Then 
\begin{gather*}
I_{i-1}+I_i=
\int_{\phi_{i-1}}^{\phi_{i+1}}\frac{u_i}{\phi_i-x} dx -d_+(\phi_{i+1}-\phi_i)-d_-(\phi_i-\phi_{i-1})=\\
=u_i\ln\frac{\phi_{i}-\phi_{i-1}}{\phi_{i+1}-\phi_i}-d_+(\phi_{i+1}-\phi_i)-d_-(\phi_i-\phi_{i-1})=
u_i\ln\frac{\phi_{i}-\phi_{i-1}}{\phi_{i+1}-\phi_i}-u_{i+1}+u_{i-1}.
\end{gather*}
\section{Eigenvalues and eigenfunctions of the modulation iteration operator}
\label{app:eval}
Here we explore properties of the transformation operator, acting on the Fourier spectrum of the modulation
according to~\eqref{eq:fiter}
\[
\hat{T} F(\w)=A(\w)F(\w)+B(\w)F(\w+2)\;,\quad \w>1\;.
\]
Here we restrict our attention to the case $\w>1$ only, because negative frequencies are symmetric to
positive ones. The real factors $A,B >0$ are given by~\eqref{eq:damp}. 

First, we employ the triangle inequality
\[
||A(\w)F(\w)+B(\w)F(\w+2)||\leq A(\w)||F(\w)||+B(\w)||F(\w+2)||\;,
\]
and restrict our attention to real positive functions $F(\w)$ only. We look for eigenfunctions
of the operator $\hat{T}$ in the class of exponentially decreasing functions of $\w$
at large frequencies: $F(\w)\sim \exp[-\alpha\w]$.
This means that we consider smooth modulations only. With an ansatz 
\[
F(\w)=f(\w)\exp[-\alpha\w] \;,
\]
where $f(\w)$ has finite variation at $\w\to\infty$,
the problem reduces to finding eigenfunctions and eigenvalues of the operator
\begin{equation}
\hat{T}_\alpha f(\w)=A(\w)f(\w)+B(\w)e^{-2\alpha} f(\w+2)\;.
\label{eq:appC1}
\end{equation}
In the simplest case of the analytic proxi-phase, where $A=B=\frac{1}{2}$, one can see that any periodic function
$f(\w)=f(\w+2)$ is an eigenfunction of~\eqref{eq:appC1} with eigenvalue 
\begin{equation}
\lambda(\alpha)=[1+\exp(-2\alpha)]/2 <1\;.
\label{eq:appC2}
\end{equation}
This shows that under action of the operator $\hat{T}$, the spectral function tends to zero exponentially, the exponent
depends on the asymptotics of the spectrum at large frequencies.

In the case of the length proxi-phase, where $A(\w)=\frac{\w-1}{2\w}$ and $B(\w)=\frac{\w+1}{2\w}$, we cannot
find the eigenfunctions, but can show that the eigenvalues are the same as for the case $A=B=1/2$. Indeed, 
the eigenvalue problem $\hat{T}_\alpha f=\lambda_\alpha f$ reduces, after substitution to \eqref{eq:appC1},
to a recursion
\[
f(\w+2)=\frac{\lambda_\alpha-A(\w)}{B(\w)e^{-2\alpha}} f(\w)\;.
\]
Because asymptotically $f(\w)$ should neither grow nor decay at $\w\to\infty$, we have
\[
\lim_{\w\to\infty}\frac{\lambda_\alpha-A(\w)}{B(\w)e^{-2\alpha}}=
\frac{\lambda_\alpha-\frac{1}{2}}{\frac{1}{2}e^{-2\alpha}}=1\;.
\]
This yields the same eigenvalue~\eqref{eq:appC2}. 

\end{appendix}

\end{document}